\documentclass[12pt,a4paper]{article}

\title {Special solutions of nonlinear von Neumann equations}
\author{  Jan Naudts$^\circ$ and Maciej Kuna$^\dagger$,\\
          \small
          $^\circ$Departement Natuurkunde, Universiteit Antwerpen,\\
          \small
          Groenenborgerlaan 171, 2020 Antwerpen, Belgium\\
          \small
          $^\dagger$Wydzia\l\ Fizyki Technicznej i Matematyki Stosowanej,\\
	  \small
	  Politechnika Gda\'{n}ska,\\
	  \small
	  ul. Narutowicza 11/12, 80-952 Gda\'{n}sk, Poland\\
	  \small
	  E-mail: jan.naudts@ua.ac.be,
	  maciek@mifgate.mif.pg.gda.pl.
}
\date{February 2006}

\usepackage{amsfonts}
\usepackage{amsmath}

\def\Io{{\mathbb I}}

\def\Tr{\,{\rm Tr}\,}
\def\sech{\,{\rm sech}}
\def\cn{\,{\rm cn}}
\def\sn{\,{\rm sn}}
\def\dn{\,{\rm dn}}

\newtheorem{lemma}{Lemma}
\newtheorem{theorem}{Theorem}
\newenvironment{proof}
{\par\noindent {\bf Proof}}
{\par\strut\hfill$\square$\par\vskip 0.5cm}

\newcommand{\be}{\begin{eqnarray}}
\newcommand{\ee}{\end{eqnarray}}

\begin{document}
\maketitle

\begin {abstract}
We consider solutions of the non-linear von Neumann equation
involving Jacobi's elliptic functions sn, cn, and dn,
and 3 linearly independent operators.
In two cases one can construct a state-dependent Hamiltonian which is such that
the corresponding non-linear von Neumann equation is solved by the given density operator.
We prove that in a certain context these two cases are the only possibilities to obtain
special solutions of this kind.
Well-known solutions of the reduced Maxwell-Bloch equations produce examples
of each of the two cases. Also known solutions of the non-linear von Neumann
equation in dimension 3 are reproduced by the present approach.
\end {abstract}

\section{Introduction}

Exact solutions of non-linear differential equations are often based on
sets of special functions like tanh and sech or Jacobi's double
periodic functions sn, cn, and dn. Here, the algebraic relations between these
functions are exploited to construct solutions of non-linear von Neumann
equations.
Given a Hamilton operator $H$, the von Neumann equation reads
\be
i\dot\rho=\bigg[H,\rho\bigg]_-.
\label {lvne}
\ee
The notations
\be
\dot\rho=\frac {{\rm d}\,}{{\rm d}t}\rho
\qquad\hbox{ and }\quad
\bigg[H,\rho\bigg]_-=H\rho-\rho H
\ee
are used. A non-linear von Neumann equation is obtained when $H$ is allowed to
depend on $\rho$
\be
i\dot\rho=\bigg[H[\rho],\rho\bigg]_-.
\label {nlvne}
\ee
The notation $H[\rho]$ is used instead of $H(\rho)$ to stress
that $ H[\rho]$ is not a function of $\rho$ in the sense of spectral theory.
Alternatively, a function $f(x)$ is given and the equation is written as $i\dot\rho=[H,f(\rho)]_-$.
The prototype of the latter, obtained with the choice $f(x)=x^2$, is the
Euler top equation \cite {LC98} and can be written as well in the form (\ref {nlvne})
\be
i\dot\rho=\bigg[H,\rho^2\bigg]_-=\bigg[\{H,\rho\},\rho\bigg]_-
\qquad\hbox{ with }\quad
\{H,\rho\}=H\rho+\rho H.
\ee
Some explicit solutions \cite {LC98,CKLN00} of this equation depend on time $t$ as $\sech(\omega t)$.
They have been called \cite {CKLN00} self-scattering solutions because they exhibit
different behaviour in the $t\rightarrow -\infty$ and $t\rightarrow +\infty$ limits
and make the transition from one behaviour to the other in a limited region of time.
However, up to now, the reason why the sech function appears
had not been investigated in a systematic manner.

It has been observed \cite {CDSW00} that solutions of the von Neumann equation with
time-dependent Hamiltonian are also solutions of a non-linear von Neumann equation.
The main application of the present work supports this point of view.
The reduced Maxwell-Bloch equations describe the time evolution of a spin variable
in presence of an applied field.
It was noted \cite {KM04} that the well-known McCall-Hahn \cite {CH67,AE87} solution of these equations
is a solution of a non-linear von Neumann equation as well. This link between time-dependent
and state-dependent Hamiltonians is clarified in the present paper.

The non-linear von Neumann equation has mainly been investigated
with Darboux transformations, trying to find non-trivial solutions
(see e.g.~\cite {LC98,CKLN00,KM04,KCL99,SWC00,UCKL01,CCU03}). Many of the solutions
constructed in this way concern 3-by-3 matrices. The connection
between these special solutions and Jacobi's double periodic
functions was noticed in the Appendix of \cite {CKLN00}.
Here, we reproduce these solutions as special cases of our approach.

In the next section, some properties of solutions of non-linear
von Neumann equations are discussed. In Section 3 we review basic
properties of Jacobi's double periodic functions and formulate methods
to construct state-dependent Hamiltonians for which the corresponding
non-linear von Neumann equations have special solutions.
Two cases are formulated and for each case a theorem is formulated
which shows that this case is the only possible solution of the non-linear von
Neumann equation that has the given form.
Section 4 treats different applications. The paper ends with a short
summary in Section 5. The proofs of the two theorems are given in Appendix.

\section {Non-linear von Neumann equation}

\subsection {Definition}

We restrict ourselves to non-linear von Neumann equations of the form (\ref {nlvne}),
with a Hamiltonian which depends linearly on the density operator $\rho$.
E.g., $ H[\rho]$ might be of the form
\be
 H[\rho]=\sum_{jk}\lambda_{jk}X_j^*\rho X_k.
\ee
The operators $X_j$ and the matrix coefficients $\lambda_{jk}$ do not depend on time.

\subsection {Spectrum}

A nice property of solutions of the non-linear von Neumann equation
is conservation of spectrum. Indeed, let be given a solution $\rho_t$ of (\ref {nlvne}), and let
$U_t$ be solution of
\be
\frac {{\rm d}\,}{{\rm d}t}U_t=iU_tH[\rho_t]
\label {unitevol}
\ee
with initial condition $U_0=\Io$. Then one has
\be
\rho_t=U^*_t\rho_0U_t.
\ee
The adjoint operators $U_t^*$ are isometric. To see this, note that
\be
\frac {{\rm d}\,}{{\rm d}t}U_tU_t^*=0
\ee
because $H[\rho_t]$ is self-adjoint. Hence one has $U_tU^*_t=\Io$,
i.e.~$U^*_t$ is isometric.
Assume now $\rho_0\psi=\lambda\psi$. Then there follows
$\rho_tU_t^*\psi=\lambda U_t^*\psi$. 
Conversely, if $\rho_t\psi=\lambda\psi$ then $\rho_0U_t\psi=\lambda U_t\psi$ follows.
Hence, $\rho_t$ and $\rho_0$ have the same spectrum.

Note that the solution of (\ref {unitevol}) can be written as
\be
U_t=\Io+i\int_0^t{\rm d}t_1\,H[\rho_{t_1}]
+i^2\int_0^t{\rm d}t_1\,\int_0^{t_1}{\rm d}t_2\,H[\rho_{t_2}]H[\rho_{t_1}]
+\cdots.
\ee

\subsection {Linear part}

The Hamiltonian $H[\rho]$ may contain a contribution of the form
$(\Tr\rho)H_0$, which for properly normalized density operators leads
to the standard linear von Neumann equations. For simplicity, this contribution
is called here the linear part. Quite often it can be
eliminated, leading to a problem which can be discussed more easily.

Let $\rho_t$ be a density operator satisfying the non-linear von Neumann
equation (\ref {nlvne}).
Let $H_0$ be a symmetric operator satisfying
for all $t$
\be
 H[e^{itH_0}\rho_t e^{-itH_0}]=e^{itH_0} H[\rho_t] e^{-itH_0}.
\label {hnotcond}
\ee
Introduce a linear map $K[\sigma]$ by
\be
K[\sigma]=H[\sigma]-(\Tr\sigma)H_0.
\label {Ksigma}
\ee
Then $\sigma_t$, defined by
\be
\sigma_t=e^{itH_0}\rho_t e^{-itH_0},
\label {rhosigma}
\ee
satifies the non-linear von Neumann equation (\ref {nlvne}) with
$H[\rho]$ replaced by $K[\sigma]$
\be
i\dot \sigma_t=\left[K[\sigma_t],\sigma_t\right]_-.
\ee

\section {Special solutions}

\subsection {Jacobi's elliptic functions}

Solutions of non-linear equations are often based on
Jacobi's elliptic functions with elliptic modulus $k$.
They satisfy the algebraic relations
\be
1&=&\sn^2(x,k)+\cn^2(x,k)\cr
1&=&\dn^2(x,k)+k^2\sn^2(k,x)\cr
\frac {{\rm d}\,}{{\rm d}x}\sn(x,k)&=&\cn(x,k)\dn(x,k)\cr
\frac {{\rm d}\,}{{\rm d}x}\cn(x,k)&=&-\sn(x,k)\dn(x,k)\cr
\frac {{\rm d}\,}{{\rm d}x}\dn(x,k)&=&-k^2\sn(x,k)\cn(x,k).
\ee
In the limit $k=1$ the period of these elliptic functions diverges
and $\sn(x,k)$ converges to $\tanh(x)$, $\cn(x,k)$ and $\dn(x,k)$ both
converge to $\sech(x)$.
The above relations then simplify to
\be
& &1=\sech^2 u+\tanh^2 u\cr
& &\frac {{\rm d}\,}{{\rm d}u}\sech u=-\sech u\tanh u\cr
& &\frac {{\rm d}\,}{{\rm d}u}\tanh u=\sech^2 u.
\ee
In the limit $k=0$ the function $\sn(x,k)$ converges to $\sin(x)$, $\cn(x,k)$ to $\cos(x)$,
and $\dn(x,k)$ converges to 1.

We next consider special solutions of the non-linear von Neumann equation involving three
independent operators. Two different cases are considered.

\subsection {Case 1}

\begin {theorem}

Let be given $\rho(t)$ of the form
\be\nonumber
\rho(t)=\theta
+\cn(\omega t,k)A+\sn(\omega t,k)B
+\dn(\omega t,k)X.
\label {case1do}
\ee

Assume $\rho(t)$ satisfies the nonlinear von Neumann equation
\be\nonumber
\dot\rho=i\left[\rho,H[\rho]\right]_-.
\label {nlvnec2}
\ee
with $H[\cdot]$ linear.

Assume that $A$, $B$, and $X$ linearly independent operators.

Assume that $\theta$ commutes with $A$, $B$, and $X$.

Assume $\omega\not=0$ and $0<k\le 1$.

Assume that the range of $H[\cdot]$ is in the linear span of the operators $A$, $B$, and $X$.

Then there exist constants $\alpha$ and $\beta$ such that
\be
i[B,X]&=&\alpha A
\label {kbx}\\
i[A,B]&=&k^2\beta X
\label {kab}\\
i[A,X]&=&-\frac {\alpha\beta}{\alpha+\beta}B.
\label {kax}
\ee
The Hamiltonian is of the form
\be
H[A]&=&\left(\nu +\frac {\omega}{\beta}\right)A\\
H[B]&=&\nu B\\
H[X]&=&\left(\nu -\frac\omega\alpha\right)X\\
H[\theta]&=&0.
\ee

Conversely, given operators $A$, $B$, $X$, and $\theta$, satisfying the above assumptions,
then the operator $\rho(t)$ given by (\ref {case1do}) solves the non-linear von Neumann equation.
\end {theorem}

One direction of the proof is given in Appendix A.
The proof of the converse statement is a straightforward exercise.

\subsection {Case 2}

\begin {theorem}

Let be given $\rho(t)$ of the form
\be\nonumber
\rho(t)=\theta
+\cn(\omega t,k)A+\sn(\omega t,k)\dn(\omega t,k)C
+\cn(\omega t,k)^2D,
\label {case2do}
\ee
with $\theta$, $A$, $C$, and $D$ linearly independent operators,
and with $\omega\not=0$ and $0<k\le 1$.

Assume $\rho(t)$ satisfies the nonlinear von Neumann equation
\be\nonumber
\dot\rho=i\left[\rho,H[\rho]\right]_-.
\ee
with $H[\cdot]$ linear.
Assume that the range of $H[\cdot]$ is in the linear span of the operators $A$, $C$, and $D$.

Assume that an operator $\theta_0$ exists, which commutes with $A$, $C$, and $D$,
such that $\theta-\theta_0$ belongs to the linear span of $A$, $C$, and $D$.

Then there exist constants $\alpha$ and $\delta$ such that
\be
i[C,D]&=&\alpha A
\label {kcd}\\
i[A,C]&=&\delta D
\label {kac}\\
i[A,D]&=&-k^2\delta C.
\label {kad}
\ee
The Hamiltonian is necessarily of the form
\be
H[A]&=&\left(\nu +\frac {2\omega}{\delta}\right)A\\
H[C]&=&\nu C\\
H[D]&=&\nu D\\
H[\theta]&=&\left(-\frac{\omega}{\alpha}+\frac {1-2k^2}{2k^2}\nu\alpha+\frac 12\delta\nu\right) D.
\label {case2ham}
\ee
The operator $\theta_0$ is given by
\be
\theta_0&=&\theta-\left(\frac {1-2k^2}{2k^2}-\frac \delta{2\alpha}\right)D.
\label {thetares}
\ee
Conversely, given operators $A$, $C$, $D$, $\theta_0$ and $\theta$, satisfying the above assumptions,
then the operator $\rho(t)$ given by (\ref {case2do}) solves the non-linear von Neumann equation.

\end {theorem}

The difficult direction of the proof is given in Appendix B.
The proof of the converse statement is a straightforward exercise.

\section {Examples}

\subsection {Reduced Maxwell-Bloch equations}

Let $\sigma_1,\sigma_2,\sigma_3$ denote the Pauli matrices.
Consider the density operator
\be
\rho_t=\frac 12(1+\sum_{\alpha=1}^3u_\alpha\sigma_\alpha)
=\frac 12\left(
\begin{array}{lcr}
1+u_3 &u_1-iu_2\\
u_1+iu_2 & 1-u_3\\
\end{array}
\right)
\ee
with
\be
u_1&=&\frac {2\tau\Delta}{1+(\tau\Delta)^2}\sech\left(t/\tau\right)\cr
u_2&=&\frac {2}{1+(\tau\Delta)^2}\sech\left(t/\tau\right)\tanh\left(t/\tau\right)\cr
u_3&=&-1+\frac {2}{1+(\tau\Delta)^2}\sech^2\left(t/\tau\right).
\label {rbsol}
\ee
See (Eq.~(4.21) of \cite {AE87}).
It solves the reduced Maxwell-Bloch equations
\be
\dot u_1&=&-\Delta u_2\cr
\dot u_2&=&\Delta u_1+\kappa{\cal E}u_3\cr
\dot u_3&=&-\kappa {\cal E}u_2
\label {rbeq}
\ee
with constants $\Delta$, $\kappa$, and $\tau$ and with
\be
{\cal E}(t)&=&\frac 2{\kappa\tau}\sech(t/\tau).
\ee

This solution satisfies the requirement that $|u|$ is constant in time
so that the eigenvalues of $\rho$, which equal $(1/2)(1\pm|u|)$,
are conserved. In fact, one has $|u|=1$. Hence, $\rho$ is a one-dimensional projection
operator.

This density operator is an example of Case 2, with $k=1$, $\omega=1/\tau$,  and with operators
\be
\theta_0&=&\frac 12\Io\\
A&=&\frac {\tau\Delta}{1+(\tau\Delta)^2}\sigma_1\\
C&=&\frac {1}{1+(\tau\Delta)^2}\sigma_2\\
D&=&\frac {1}{1+(\tau\Delta)^2}\sigma_3.
\ee
They satisfy the commutation relations
\be
i[A,C]_-&=&-\frac {2\tau\Delta}{1+(\tau\Delta)^2}D\\
i[C,D]_-&=&-\frac 2{\tau\Delta}\frac {1}{1+(\tau\Delta)^2}A\\
i[D,A]_-&=&-\frac {2\tau\Delta}{1+(\tau\Delta)^2}C.
\ee

From Theorem 2 follows that $\rho_t$ satisfies the non-linear von Neumann equation
with Hamiltonian given by
\be
H[\sigma_1]&=&-\frac 1\Delta(\omega^2+\Delta^2)\sigma_1\\
H[\sigma_2]&=&0\\
H[\sigma_3]&=&0\\
H[\Io]&=&-\frac {2\omega^3}{\omega^2+\Delta^2}\sigma_3.
\ee

\subsection {Including phase modulation}

A related problem is obtained by including phase modulation effects.
The equations are
\be
\dot u_1&=&\dot\phi u_2\cr
\dot u_2&=&-\dot\phi u_1+\kappa{\cal E}u_3\cr
\dot u_3&=&-\kappa{\cal E}u_2
\ee
with
\be
\dot\phi&=&-\delta\tanh(t/\tau)\cr
{\cal E}&=&\frac 1{\kappa\tau}\sqrt{1+(\tau\delta)^2}\sech(t/\tau).
\ee
Solutions are (Eq.~(4.49) of \cite {AE87})
\be
u_1&=& \tau\delta u_2\cr
u_2&=&-\frac 1{\sqrt{1+(\tau\delta)^2}}\sech(t/\tau)\cr
u_3&=&\tanh(t/\tau).
\label {phasmodsol}
\ee
Note that
\be
\dot u_1&=&-\frac 1\tau u_1u_3\cr
\dot u_2&=&-\frac 1\tau u_2u_3\cr
\dot u_3&=&\frac 1\tau (1-u_3^2)=\frac 1\tau(u_1^2+u_2^2).
\ee

Let us try to match this example to Case 1. Clearly needed is $\omega=1/\tau$, $k^2=1$,
$\theta=(1/2)\Io$, and $B=(1/2)\sigma_3$. The choice of operators $A$ and $X$ is not so obvious.
But the choice
\be
A&=&-\frac 12\frac {\tau\delta}{\sqrt{1+\tau^2\delta^2}}\sigma_1\\
X&=&-\frac 12\frac {1}{\sqrt{1+\tau^2\delta^2}}\sigma_2\\
\ee
satisfies the requirements.
From Theorem 1 then follows that $\rho_t$ satisfies the non-linear von Neumann equation with
Hamiltonian given by
\be
H[\sigma_1]&=&\frac{\omega^2}{\delta}\sigma_1\\
H[\sigma_2]&=&-\delta\sigma_2\\
H[\sigma_3]&=&0\\
H[\Io]&=&0.
\ee

\subsection {Three-level system}

Fix $0<k\le 1$ and $\alpha$, $\delta$, and $\phi$ real.
Consider 3-by-3 matrices
\be
A &=& k\frac{\delta}{\sqrt{2}}
\left(\begin {array}{lcr}
0 &0 &e^{i\phi}\\
0 &0 &e^{i\phi}\\
e^{-i\phi} &e^{-i\phi} &0
\end {array}\right),\\
C &=& \sqrt{\frac{\alpha\delta}{2}}
\left(\begin {array}{lcr}
0 &0 &-ie^{i\phi}\\
0 &0 &ie^{i\phi}\\
ie^{-i\phi} &-ie^{-i\phi} &0
\end {array}\right),\\
D &=& k \sqrt{\alpha\delta}
\left(\begin {array}{lcr}
1 &0 &0\\
0 &-1 &0\\
0 &0 &0
\end {array}\right),\\
\theta_0 &=& \frac{1}{3} \mathbb{I}.
\ee
It is easy to verify that they satisfy the case 2 commutation relations (\ref {kcd}, \ref {kac}, \ref {kad}).
Hence the density matrix $\rho(t)$, given by (\ref {case2do}), satisfies the non-linear von Neumann equation 
(\ref {nlvnec2}) for any Hamiltonian of the form (\ref {case2ham}).
The operator $\theta_0$ is given by (\ref {thetares}).

This result can be transformed by adding a linear part (in the sense of Section 2.3).
Let $\sigma (t) = e^{-i\mu tP_3}\rho(t)e^{i\mu tP_3}$ with
\be
P_3= \left(\begin {array}{lcr}
0 &0 &0\\
0 &0 &0\\
0 &0 &1
\end {array}\right).
\ee
Note that
\be
H[\sigma(t)]=e^{-i\mu tP_3}H[\rho(t)]e^{i\mu tP_3}
\ee
is trivially satisfied because $H[A]$ is proportional to $A$,
$H[C]$ is proportional to $C$, $H[D]$ is proportional to $D$, and $\theta$ commutes with $P_3$.
This implies that $\sigma(t)$ satisfies the equation
\be
i\dot{\sigma} = \left[H[\sigma] + \mu P_3,\sigma\right]
\label {MB}
\ee
The latter equation can be written into the form
\be
 i\dot{\sigma} = \left[H_0 + H_I,\sigma\right],
 \ee
where
 \be
 H_0 = \left(\begin {array}{lcr}
\lambda &0 &0\\
0 &-\lambda &0\\
0 &0 &\mu
\end {array}\right),
\ee
and
\be
 H_I =  \epsilon \cn(\omega t,k)\left(\begin {array}{lcr}
0 &0 &e^{i(\phi - \mu t)}\\
0 &0 &e^{i(\phi -\mu t)}\\
e^{-i(\phi -\mu t)} &e^{-i(\phi - \mu t)} &0
\end {array}\right),
\ee
and $\lambda = -k\omega\sqrt{\frac{\delta}{\alpha}}$,
$\epsilon = \frac{k\omega}{\sqrt{2}}$, $\nu=0$. It describes a three-level system
interacting with an electromagnetic pulse $E =E_0 e^{i(\phi - \mu t)}\cn(\omega t,k)$.

\subsection {Known solutions in $d=3$}

Fix real constants $k$, $\omega$, $\phi$, $\lambda$ and $\mu$, satisfying $|\lambda|<\mu$ and $0<k\le 1$.
Define operators $\theta$, $A$, $B$, and $X$, by
\be
\theta &=& \frac{1}{3} \Io,\\
B &=& \frac{k\omega}{\sqrt{{\mu}^2 - {\lambda}^2}} \left(\begin {array}{lcr}
0 &1 &0\\
1 &0 &0\\
0 &0 &0
\end {array}\right)\\
A &=& \frac{k\omega}{\sqrt{2{\mu}(\mu + {\lambda})}} \left(\begin {array}{lcr}
0 &0 &e^{i\phi}\\
0 &0 &0\\
e^{-i\phi} &0 &0
\end {array}\right)\\
X &=& \frac{\omega}{\sqrt{2{\mu}(\mu - {\lambda})}} \left(\begin {array}{lcr}
0 &0 &0\\
0 &0 &-ie^{i\phi}\\
0 &ie^{-i\phi} &0
\end {array}\right).
\ee
They satisfy the case-1 commutation relations with
\be
\alpha&=&\frac {\omega}{\mu-\lambda}\\
\beta&=&\frac {\omega}{\mu+\lambda}.
\ee
Hence, the results of the previous section imply that the density operator
\be
\sigma_t=\theta+B\sn(\omega t,k)+A\cn(\omega t,k)+X\dn(\omega t,k)
\ee
satisfies the non-linear von Neumann equation
$i\dot\sigma=[H[\sigma],\sigma]_-$ for any Hamiltonian $H[\sigma]$
satisfying
\be
H[\theta]&=&0\\
H[A]&=&(\nu+\mu+\lambda)A\\
H[B]&=&\nu B\\
H[X]&=&(\nu-\mu+\lambda)X.
\ee
A satisfactory choice is
\be
H[\sigma]&=&
\mu\left(\begin{array}{lcr}
0 &0 &\sigma_{31}\\
0 &0 &-\sigma_{32}\\
\sigma_{13} &-\sigma_{23} &0
\end{array}\right)
+\lambda\left(\begin{array}{lcr}
0 &0 &\sigma_{31}\\
0 &0 &\sigma_{32}\\
\sigma_{13} &\sigma_{23} &0
\end{array}\right).
\ee
Note that
\be
H[\sigma]&=&\{H_0,\sigma\}-\frac 23(\Tr\sigma)H_0
\ee
with
\be
H_0&=&\left(\begin {array}{lcr}
\mu &0 &0\\
0 &-\mu &0\\
0 &0 &\lambda
\end{array}\right).
\label {cklnham}
\ee
In particular, $\rho_t$ defined by
\be
\rho_t=e^{-(2/3)it H_0}\sigma_te^{(2/3)it H_0}
\ee
satisfies the non-linear von Neumann equation $i\dot\rho=[H_0,\rho^2]_-$.
This equation with Hamiltonian (\ref {cklnham}) has been studied in \cite {CKLN00}. Explicit solutions
were obtained in the limit $k=1$.

\subsection {Variations on a theme}

Let us slightly modify the previous example. Fix real constants $b\not=0$,
$\omega$, $\phi$, and $0<k\le 1$. 
Define operators $\theta$, $A$, $B$, and $X$, by
\be
\theta &=& \frac{1}{3} \Io,\\
A &=& \frac{k\omega}{b\sqrt{2}} \left(\begin {array}{lcr}
0 &1 &0\\
1 &0 &0\\
0 &0 &0
\end {array}\right)\\
B &=& \frac{k\omega}{b} \left(\begin {array}{lcr}
0 &0 &e^{i\phi}\\
0 &0 &0\\
e^{-i\phi} &0 &0
\end {array}\right)\\
X &=& \frac{\omega}{b\sqrt{2}} \left(\begin {array}{lcr}
0 &0 &0\\
0 &0 &-ie^{i\phi}\\
0 &ie^{-i\phi} &0
\end {array}\right).
\ee
They satisfy the case-1 commutation relations with
\be
\alpha=\beta=-\frac {\omega}b.
\ee
The  density operator
\be
\sigma_t=\theta+B\sn(\omega t,k)+A\cn(\omega t,k)+X\dn(\omega t,k)
\ee
satisfies the non-linear von Neumann equation
$i\dot\sigma=[H[\sigma],\sigma]_-$ for any Hamiltonian $H[\sigma]$
satisfying
\be
H[\theta]&=&0\\
H[A]&=&(\nu-b)A\\
H[B]&=&\nu B\\
H[X]&=&(\nu+b)X.
\ee
A satisfactory choice, corresponding with $\nu=4b$, is
\be
H[\sigma]&=&
b\left(\begin{array}{lcr}
0 &3\sigma_{21} &4\sigma_{31}\\
3\sigma_{13} &0 &5\sigma_{32}\\
4\sigma_{13} &5\sigma_{23} &0
\end{array}\right).
\ee
With this choice is
\be
H[\sigma]&=&\{H_0,\sigma\}-\frac 23(\Tr\sigma)H_0,
\ee
with
\be
H_0&=&b\left(\begin {array}{lcr}
1 &0 &0\\
0 &2 &0\\
0 &0 &3
\end{array}\right).
\ee
In particular, $\rho_t$ defined by
\be
\rho_t=e^{-(2/3)it H_0}\sigma_te^{(2/3)it H_0}
\ee
satisfies the non-linear von Neumann equation $i\dot\rho=[H_0,\rho^2]_-$.
In the limit $k=1$ this example has been discussed in \cite {KM04}.

\section {Discussion}

This paper studies the non-linear von Neumann equation under the restrictions that
(1) the Hamiltonian $H[\rho]$ depends linearly on the density operator $\rho$;
(2) the solution $\rho_t$ involves Jacobi's elliptic functions sn, cn and dn,
or there limits tanh and sech;
(3) the solution $\rho_t$ involves 3 linearly independent operators, and possibly a fourth
operator commuting with these three operators.

The paper does not focus on methods for solving non-linear von Neumann equations.
It rather follows the opposite way. Starting from a special solution $\rho_t$, it constructs the Hamiltonian
$H[\rho]$ for which $\rho_t$ solves the corresponding non-linear von Neumann equation.
Two cases have been treated.
In both cases we have proved a theorem stating conditions under which the special solution is found.
Two well-known solutions of the reduced Maxwell-Bloch equations are examples of
these two cases.
We have shown that the reduced Maxwell-Bloch equations can be generalized to three-level systems
and that these have not only special solutions involving sech and tanh, but also periodic
solutions involving  Jacobi's elliptic functions.
Finally, some of the known 3-dimensional solutions of
the non-linear von Neumann equation appear to be examples of these two cases as well.

The examples of the reduced Maxwell-Bloch equations show that non-linear
von Neumann equations and their solutions appear in physics in a natural
manner. The $d=3$-solutions of the equation $i\dot\rho=[H_0,\rho^2]$,
obtained here, are generalizations of those found in \cite {KM04}.

\section* {Acknowledgements}

We thank Marek Czachor ans Sergiej Leble for interesting discussions on the
reduced Maxwell-Bloch equations.
This work has been supported by the Flemish-Polish bilateral project
"Soliton techniques applied to equations of quantum field theory"
and by the project "Soliton concept in classical and quantum contexts"
of the Flemish Fund for Scientific Research FWO.

\appendix
\section {Proof of Theorem 1}

A straightforward calculation leads to the set of equations
\be
0&=&i[H[A],A]_-+i[H[X],X]_-\label {c1eq1}\\
0&=&-i[H[A],A]_-+i[H[B],B]_--k^2i[H[X],X]_-\label {c1eq2}\\
0&=&i[H[\theta],A]_-\label {c1eq3}\\
0&=&i[H[\theta],B]_-\label {c1eq4}\\
0&=&i[H[\theta],X]_-\label {c1eq5}\\
\omega A&=&i[H[B],X]_-+i[H[X],B]_-\label {c1eq6}\\
-\omega B&=&i[H[A],X]_-+i[H[X],A]_-\label {c1eq7}\\
k^2\omega X&=&i[H[A],B]_-+i[H[B],A]_-.\label {c1eq8}
\ee
Equations (\ref {c1eq3}, \ref {c1eq4}, \ref {c1eq5}) imply that $H[\theta]$
commutes with $A$, $B$, and $X$.

Introduce the notations
\be
H[A]&=&a_AA+b_AB+x_AX\\
H[B]&=&a_BA+b_BB+x_BX\\
H[X]&=&a_XA+b_XB+x_XX\\
H[\theta]&=&a_0A+b_0B+x_0X.
\ee
Then the remaining equations become
\be
0&=&-b_AK_{AB}+(a_X-x_A)K_{AX}+b_XK_{BX}\\
0&=&(a_B+b_A)K_{AB}+(x_A-k^2a_X)K_{AX}-(x_B+k^2b_X)K_{BX}\\
0&=&-b_0K_{AB}-x_0K_{BX}\\
0&=&a_0K_{AB}-x_0K_{BX}\\
0&=&a_0K_{AX}+b_0K_{BX}\\
\omega A&=&a_XK_{AB}+a_BK_{AX}+(b_B-x_X)K_{BX}\\
-\omega B&=&-b_XK_{AB}+(a_A-x_X)K_{AX}+b_AK_{BX}\\
k^2\omega X&=&(a_A-b_B)K_{AB}-x_BK_{AX}-x_AK_{BX}.
\ee
Because $A$, $B$, and $X$ are linearly independent and $\omega$ and $k$ do not vanish
the last three equations imply that $K_{AB}$, $K_{AX}$, and $K_{BX}$ are linearly independent.
Then the first five equations imply $a_B=a_X=b_A=b_X=x_A=x_B=a_0=b_0=c_0=0$.
The three last equations then read
\be
\omega A&=&(b_B-x_X)K_{BX}\\
-\omega B&=&(a_A-x_X)K_{AX}\\
k^2\omega X&=&(a_A-b_B)K_{AB}.
\ee
This implies (\ref {kbx}, \ref {kab}, \ref {kax}) with
\be
\alpha&=&\frac \omega{b_B-x_X}\\
\beta&=&\frac \omega {aA-b_B}.
\ee
Let $\nu=b_B$. Then one finds
\be
a_A&=&\nu+\frac \omega\beta\\
x_X&=&\nu-\frac\omega\alpha.
\ee
The equations for $H[\cdot]$ then follow.

\section {Proof of Theorem 2}

\subsection {Equations}

A straightforward calculation leads to the set of equations
\be
0&=&i[D,H[D]]-k^2i[C,H[C]]\label {C21}\\
0&=&i[\theta,H[\theta]]+(1-k^2)i[C,H[C]]\label {C22}\\
0&=&i[\theta,H[D]]+i[D,H[\theta]]+i[A,H[A]]+(2k^2-1)i[C,H[C]]\label {C23}\\
0&=&i[C,H[D]]+i[D,H[C]]\label {C24}\\
-\omega A&=&i[\theta,H[C]]+i[C,H[\theta]]\label {C25}\\
2k^2\omega C&=&i[A,H[D]]+i[D,H[A]]\label {C26}\\
\omega C&=&i[\theta,H[A]]+i[A,H[\theta]]+i[A,H[D]]+i[D,H[A]]\label {C27}\\
-2\omega D&=&i[A,H[C]]+i[C,H[A]]\label {C28}.
\ee

Introduce the notations $K_{AB}=i[A,B]_-$, and similar notation for other commutators.
Because the range of $H[\cdot]$ is in the span of $A$, $C$, and $D$, one can write
\be
H[A]&=&a_AA+c_AC+d_AD\\
H[C]&=&a_CA+c_CC+d_CD\\
H[D]&=&a_DA+c_DC+d_DD\\
H[\theta]&=&a_0A+c_0C+d_0D.
\ee
By assumption there exist numbers $t_A$, $t_C$, and $t_D$, and an operator $\theta_0$
commuting with $A$, $C$, and $D$, such that
\be
\theta=\theta_0+t_AA+t_CC+t_DD.
\ee
This implies
\be
K_{A\theta}&=&t_CK_{AC}+t_DK_{AD}\\
K_{C\theta}&=&-t_AK_{AC}+t_DK_{CD}\\
K_{D\theta}&=&-t_AK_{AD}-t_CK_{CD}.
\ee
Then the equations (\ref {C25}, \ref {C26},\ref {C28}) become
\be
-\omega A&=&(c_Ct_A-a_Ct_C-a_0)K_{AC}+(d_Ct_A-a_Ct_D)K_{AD}\cr
& &+(d_Ct_C-c_Ct_D+d_0)K_{CD}
\label {C25a}\\
2k^2\omega C&=&c_DK_{AC}+(d_D-a_A)K_{AD}-c_AK_{CD}
\label {C26a}\\
-2\omega D&=&(c_C-a_A)K_{AC}+d_CK_{AD}+d_AK_{CD}.
\label {C28a}
\ee
Because $A$, $C$ and $D$, are linearly independent and $k$ and $\omega$ do not vanish one
concludes that the operators $K_{AC}$, $K_{AD}$, and $K_{CD}$, are linearly independent.

Consider next (\ref {C21}, \ref {C24}). They can be written as
\be
0&=&-(c_D+k^2d_C)K_{CD}-a_DK_{AD}+k^2a_CK_{AC}
\label {C21a}\\
0&=&(d_D-c_C)K_{CD}-a_CK_{AD}-a_DK_{AC}.
\label {C24a}
\ee
Because of the linear independence of $K_{AC}$, $K_{AD}$, and $K_{CD}$, 
they imply $a_C=a_D=0$ and $c_D=-k^2d_C$ and $d_D=c_C$. The equations (\ref {C25a}, \ref {C26a}, \ref {C28a})
can therefore be written as
\be
-\omega A&=&(c_Ct_A-a_0)K_{AC}+d_Ct_AK_{AD}\cr
& &+(d_Ct_C-c_Ct_D+d_0)K_{CD}
\label {C25b}\\
2k^2\omega C&=&-k^2d_CK_{AC}+(c_C-a_A)K_{AD}-c_AK_{CD}
\label {C26b}\\
-2\omega D&=&(c_C-a_A)K_{AC}+d_CK_{AD}+d_AK_{CD}.
\label {C28b}
\ee

Finally, (\ref {C22}), (\ref {C23}), and (\ref {C27}), the latter simplified with (\ref {C26}), become
\be
0&=&(c_0t_A-a_0t_C)K_{AC}+(d_0t_A-a_0t_D)K_{AD}\cr
& &+(d_0t_C-c_0t_D+(1-k^2)d_C)K_{CD}\\
0&=&(c_A-k^2d_Ct_A)K_{AC}+(c_Ct_A+d_A-a_0)K_{AD}\cr
& &+(k^2d_Ct_D+c_Ct_C+(2k^2-1)d_C-c_0)K_{CD}\\
0&=&(-k^2(1-2k^2)d_C-2k^2(c_0-a_At_C+c_At_A))K_{AC}\cr
& &+((1-2k^2)(c_C-a_A)-2k^2(d_0-a_At_D+d_At_A))K_{AD}\cr
& &+(-(1-2k^2)c_A+2k^2(c_At_D-d_At_C))K_{CD}.
\ee
Because of the independence of $K_{AC}$, $K_{AD}$, and $K_{CD}$, nine equations follow
\be
0&=&c_0t_A-a_0t_C\label {n36}\\
0&=&d_0t_A-a_0t_D\label {n37}\\
0&=&d_0t_C-c_0t_D+(1-k^2)d_C\label {n38}\\
0&=&c_A-k^2d_Ct_A\label {n39}\\
0&=&c_Ct_A+d_A-a_0\label {n40}\\
0&=&k^2d_Ct_D+c_Ct_C+(2k^2-1)d_C-c_0\label {n41}\\
0&=&-k^2(1-2k^2)d_C-2k^2(c_0-a_At_C+c_At_A)\label {n42}\\
0&=&(1-2k^2)(c_C-a_A)-2k^2(d_0-a_At_D+d_At_A)\label {n43}\\
0&=&-(1-2k^2)c_A+2k^2(c_At_D-d_At_C).\label {n44}
\ee
These equations fix $a_0$, $c_0$, $d_0$, $c_A$, and $d_C$ in terms of the remaining
parameters. The latter are constraint by another 4 equations. Their analysis is
done in the next subsection.

\subsection {Analysis}

\begin {lemma}
$c_A=0$.
\end {lemma}

\begin {proof}
Multiply (\ref {n38}) to obtain
\be
(1-k^2)a_0d_C=a_0c_0t_D-a_0d_0t_c.
\ee
Then (\ref {n36}, \ref {n37}) imply that the r.h.s.~vanishes. One concludes that $a_0d_C=0$.
This leaves two cases
\begin {description}
\item {1)} $d_C=0$; the lemma follows from (\ref {n39}).
\item {2)} $d_C\not=0$ and $a_0=0$; then (\ref {n38}) implies that $c_0$ and $d_0$ cannot both vanish.
But this implies $t_A=0$ via (\ref {n36}, \ref {n37}). Again $c_A=0$ follows from (\ref {n39}).
\end {description}
\end {proof}

\begin {lemma}\label {twocases}
Either one of the following cases holds
\begin {itemize}
\item {a)} $d_A=0$ and $a_0d_C=d_Ct_A=0$;
\item {b)} $d_A\not=0$ and $t_C=d_C=c_0=0$.
\end {itemize}
\end {lemma}

\begin {proof}
From (\ref {n44}), using $c_A=0$, follows $d_At_C=0$. Hence either $d_A=0$ or $d_A\not=0$ and $t_C=0$.

Use that $a_0d_C=d_Ct_A=0$ from the proof of the previous lemma. Then (\ref {n40}), multiplied with $d_C$,
implies that $d_Ad_C=0$. Hence the assumption that $d_A\not=0$ implies $d_C=0$. But $d_A\not=0$ implies also $t_C=0$.
Finally, (\ref {n41}) implies $c_0=0$.

Next assume $d_A=0$. The results $a_0d_C=d_Ct_A=0$ are taken from the proof of the previous lemma.

\end {proof}

The first possibility of the previous lemma is the only one that can occur.
This is proved in following lemma.

\begin {lemma}
 $d_A=0$.
\end {lemma}

\begin {proof}
The proof goes {\sl ex absurdo}.
Assume that the second case of the previous lemma holds, i.e.~$d_A\not=0$ and $c_A=t_C=d_C=c_0=0$.
Then the operator expressions (\ref {C25b}, \ref {C26b}, \ref {C28b}) become
\be
-\omega A&=&(c_Ct_A-a_0)K_{AC}+(-c_Ct_D+d_0)K_{CD}\\
2k^2\omega C&=&(c_C-a_A)K_{AD}\\
-2\omega D&=&(c_C-a_A)K_{AC}+d_AK_{CD}.
\ee
The solution of this set of equations is
\be
K_{AC}&=&\frac 1N\left[-d_A\omega A+2(d_0-c_Ct_D)\omega D\right]\\
K_{AD}&=&\frac {2k^2}{c_C-a_A}\omega C\\
K_{CD}&=&\frac 1N\left[(c_C-a_A)\omega A-2(c_Ct_A-a_0)\omega D\right],
\ee
with
\be
N=d_A(c_Ct_A-a_0)-(c_C-a_A)(d_0-c_Ct_D).
\ee
The Jacobi identity then implies
\be
0&=&i[A,K_{CD}]-i[C,K_{AD}]+i[D,K_{AC}]\cr
&=&\frac 1N\left[-2(c_Ct_A-a_0)+d_A\right]\omega K_{AD}.
\ee
Hence there follows $d_A=2(c_Ct_A-a_0)$. Comparison with (\ref {n40}) then implies
$d_A=0$, which contradicts the assumption that $d_A\not=0$.

\end {proof}

\begin {lemma}
$d_C=0$.
\end {lemma}

\begin {proof}
From Lemma \ref {twocases} follows $d_Ct_A=0$. This implies $d_C=0$ or $t_A=0$.
Let us assume $t_A=0$. Then, using the results of the previous lemmas, (\ref {n36} --- \ref {n44}) reduce to
\be
0&=&d_0t_C-c_0t_D+(1-k^2)d_C\label {p38}\\
0&=&a_0\label {p40}\\
0&=&k^2d_Ct_D+c_Ct_C+(2k^2-1)d_C-c_0\label {p41}\\
0&=&(1-2k^2)d_C+2(c_0-a_At_C)\label {p42}\\
0&=&(1-2k^2)(c_C-a_A)-2k^2(d_0-a_At_D)\label {p43}.
\ee
Use the last two equations to eliminate $c_0$ and $d_0$ from the first and the third equation.
This gives
\be
0&=&(1-2k^2)\left[k^2d_Ct_D+(c_C-a_A)t_C\right]+2k^2(1-k^2)d_C\\
0&=&k^2d_Ct_D+(c_C-a_A)t_C+\frac 12(2k^2-1)d_C.
\ee
Combining these two equations gives $d_C=0$.

\end {proof}

\begin {lemma}
$c_0=t_C=0$.
\end {lemma}

\begin {proof}
From (\ref {C26b}), using $c_A=d_C=0$, follows $c_C\not=a_A$.
But from (\ref {n41}, \ref {n42}) follows
\be
0=(c_C-a_A)t_C.
\ee
Hence, $t_C=0$ follows. From (\ref {n41}) or (\ref {n42}) then follows $c_0=0$.
\end {proof}

\begin {lemma}
$a_0=t_A=0$.
\end {lemma}

\begin {proof}
From (\ref {n37}, \ref {n40}) follows
\be
0=(d_0-c_Ct_D)t_A.
\ee
Note that $d_0\not=c_Ct_D$ follows from (\ref {C25b}).
One concludes therefore that $t_A=0$.

Finally, $d_A=0$ and (\ref {n40}) imply $c_Ct_A-a_0=0$.
Hence $t_A=0$ implies $a_0=0$.

\end {proof}

\subsection {Completing the proof}

Thus far, the desired form of the operators in (\ref {kcd} --- \ref {thetares}) has been proved.
The actual relations between the coefficients follow by inserting these expressions
into the equations (\ref {C21}---\ref {C28}).
In particular, the operator expressions (\ref {C25b}, \ref {C26b}, \ref {C28b}) become
\be
2k^2\omega C&=&(c_C-a_A)K_{AD}
\label {final1u1}\\
-2\omega D&=&(c_C-a_A)K_{AC}
\label {final2u1}\\
\omega A&=&(c_Ct_D-d_0)K_{CD}
\label {final5u1}.
\ee
The solution of this set of equations is of the form (\ref {kcd}, \ref {kac}, \ref {kad})
with
\be
\delta&=&-\frac {2\omega}{c_C-a_A}\\
\alpha&=&\frac {\omega}{c_Ct_D-d_0}.
\ee


\end {document}